\documentclass[%
prd,
reprint,
superscriptaddress,
nofootinbib,
amsmath,
amssymb,
aps,
]{revtex4-2}

\usepackage{booktabs}	        
\usepackage{array}		        
\usepackage[inline]{enumitem}	
\usepackage{hyperref}
\hypersetup{
    colorlinks = true,
	citecolor = blue,
	linkcolor = blue
}
\usepackage{graphicx}
\usepackage{bbm}
\usepackage{mathtools}

\newcommand{\mj}{\mathcal{J}}
\newcommand{\mf}{\mathcal{F}}
\newcommand{\tdr}{T_{\rm coh}}
\newcommand{\td}{\textrm{d}}

\begin{document}

\title{How much spin wandering can continuous gravitational wave \texorpdfstring{\\}{} search algorithms handle?}

\author{Julian B. Carlin}
\affiliation{University of Melbourne, School of Physics, Parkville, VIC 3010, Australia}
\affiliation{Australian Research Council Centre of Excellence for Gravitational Wave Discovery (OzGrav), University of Melbourne, Parkville, VIC 3010, Australia}
\author{Andrew Melatos}
\affiliation{University of Melbourne, School of Physics, Parkville, VIC 3010, Australia}
\affiliation{Australian Research Council Centre of Excellence for Gravitational Wave Discovery (OzGrav), University of Melbourne, Parkville, VIC 3010, Australia}

\date{\today}

\begin{abstract}
    The canonical signal model in continuous gravitational wave searches is deterministic, and stable over the long integration times needed to separate a putative signal from the noise, e.g.~with a matched filter. However, there exist plausible physical mechanisms that give rise to ``spin-wandering'', i.e.~small stochastic variations in the frequency of the gravitational wave. Stochastic variations degrade the sensitivity of matched filters which assume a deterministic frequency evolution. Suites of synthetic spin-wandering injections are performed to infer the loss in sensitivity depth $\mathcal{D}_{\rm SW}$ when compared to the depth for a canonical signal $\mathcal{D}_{\rm det}$. For a fiducial spin-wandering signal that wanders by $\lesssim5 \times 10^{-6}\,$Hz per day, the depth ratio is $\mathcal{D}_{\rm det} / \mathcal{D}_{\rm SW}=4.39^{+0.23}_{-0.27}$, $1.51^{+0.02}_{-0.03}$, $1.75^{+0.04}_{-0.04}$, and $1.07^{+0.01}_{-0.02}$ for the coherent $\mf$-statistic, semi-coherent $\mf$-statistic, CrossCorr, and HMM-Viterbi algorithms respectively. Increasing the coherence time of the semi-coherent algorithms does not necessarily increase their sensitivity to spin-wandering signals.
\end{abstract}

\maketitle

\section{Introduction} \label{sec:intro}
Despite the now-routine detection by the Laser Interferometer Gravitational-Wave Observatory (LIGO) \citep{aligo2015}, Virgo \citep{avirgo2014}, and Kamioka Gravitational Wave Detector (KAGRA) \citep{Akutsu2021} Collaboration (LVK) of gravitational waves from compact binary coalescences \citep{gwtc3}, continuous gravitational waves have not been detected yet. By continuous gravitational waves, we refer to long-lived ($\gtrsim 1\,$yr) quasi-monochromatic signals. The prototypical sources of such signals are rapidly rotating neutron stars with a time-varying mass (or mass current) quadrupole \citep{Misner1973,Andersson1998}. They are thought to emit at or near simple rational multiples of the star's rotation frequency \citep{JKS98,Andersson1998,Caride2019}. While the rotation frequency of neutron stars is typically stable, so-called ``spin-wandering'', i.e. small stochastic variations in frequency, may arise from accretion torque fluctuations \citep{Romanova2016,Mukherjee2018}, superfluid turbulence \citep{Melatos2010,Link2012}, or an unknown physical mechanism perhaps linked to the phenomenon of timing noise \citep{Cordes1985,Shannon2010}.

Spin-wandering limits the sensitivity of traditional search algorithms, which assume a deterministic frequency evolution. One enduring target for continuous gravitational wave searches is the low-mass X-ray binary Scorpius X-1, the brightest extra-solar X-ray source in the sky \citep{o1crosscorSco,o2vitsco,Zhang2021,o3vitsco,o3crosscorSco,Whelan2023,o3vitscoredo}. Scorpius X-1 is a potentially loud emitter of continuous gravitational waves, as the back-reaction torque from gravitational wave emission must be large if it is the mechanism which balances the accretion torque, such that the neutron star does not spin up to the break-up rotation frequency \citep{Bildsten1998}. Studies of the X-ray flux variability from Scorpius X-1 suggest that the frequency may drift by up to 50$\,\mu$Hz per year \citep{Mukherjee2018}, although this depends on the details of the accretion physics. 

Quantifying the impact of spin wandering on the performance of continuous wave search algorithms is important. A successful detection of a continuous wave source even in the absence of spin wandering would be a major scientific advance. It would elucidate the neutron star equation of state, questions of nuclear physics, superfluidity, superconductivity, and general relativity, among other things; see Refs. \citep{Glampedakis2018,Jones2022a,Riles2023,Wette2023} for recent reviews. If, in addition, the algorithm responsible for the successful detection tracks the wandering of the spin frequency, it would deliver valuable extra information about the accretion physics \citep{Melatos2023,OLeary2024}, superfluid--crust coupling \citep{Meyers2021,Meyers2021a}, or the mechanism behind timing noise \citep{ONeill2024}.

Continuous gravitational waves are hard to detect even when the assumed signal model is deterministic. The computational cost of a search using a coherent matched filter such as the $\mf$-statistic \citep{JKS98} grows with $T^n$, where the coherent integration time is $T$, and $n \geq 5$ grows rapidly with the number of terms in the Taylor expansion of the phase evolution \citep{Wette2018}. This has encouraged a wide variety of algorithmic approaches to searches, depending on the target, available computational resources, and assumed signal model. In this work we assess four mature algorithms: the coherent $\mf$-statistic \citep{JKS98,Cutler2005,Prix2007,Prix2009a}, the semi-coherent $\mf$-statistic \citep{Brady2000,Prix2012,Wette2015}, the cross-correlation algorithm (CrossCorr) \citep{crosscorInit,Whelan2015}, and the hidden Markov model solved by the Viterbi algorithm (HMM-Viterbi) \citep{Viterbi1967,Suvorova2016,Suvorova2017,Melatos2021}. The semi-coherent $\mf$-statistic computes the coherent $\mf$-statistic in $N$ disjoint segments of length $\tdr$, with no assumption of phase continuity between segments, reducing the computational cost scaling to $N^m \tdr^n$, with $m\approx 2$. CrossCorr constructs a detection statistic from a sum of products of pairs of short Fourier transforms (SFTs) separated by at most $\tdr$, which are weighted according to an assumed signal model. HMM-Viterbi models the gravitational wave frequency as a stochastically wandering hidden variable to be inferred. Some of the above algorithms competed to detect a set of blind injections in the Scorpius X-1 Mock Data Challenge \citep{scox1mdc1,Suvorova2017}. The goal of the MDC was to compare the efficacy of algorithms when faced with identical sets of injections, but all signals were deterministic. 

In this paper we investigate empirically the efficacy with which the above four continuous gravitational wave search algorithms can detect synthetic spin-wandering signals. In Section~\ref{sec:signalmodel} we introduce both the canonical and spin-wandering signal models. In Section~\ref{sec:inj} we describe the suites of synthetic signal injections we perform. Section~\ref{sec:alg} briefly outlines the pertinent algorithmic parameter choices we make when running the search algorithms. In Section~\ref{sec:det_eff} we calculate the sensitivity depths achieved by the search algorithms when faced with spin-wandering signals. We conclude in Section~\ref{sec:concl}.

\section{Signal model} \label{sec:signalmodel}
Different search algorithms assume different models for the evolution of the signal phase during an observation. In this section, we define the deterministic (equivalently ``canonical'') phase model used by most continuous gravitational wave search algorithms, including the coherent $\mf$-statistic, the semi-coherent $\mf$-statistic, and CrossCorr (Section~\ref{sec:detphase}) and the stochastic signal model extension used by searches with the HMM-Viterbi algorithm (Section~\ref{sec:sw_model}). It is important to note that the sensitivities of algorithms based on different phase models usually cannot be compared directly; for example, upper limits inferred from nondetections are conditional and calibrated with synthetic injections which implicitly encode the assumed signal model. This point is recognized when presenting results from published searches to date \citep{Beniwal2021,o3aSNR,o3amxp,o3vitsco,Wette2023,o3vitscoredo}.

\subsection{Deterministic phase} \label{sec:detphase}
The canonical evolution of the emission frequency $f_{\rm GW}$ of a continuous gravitational wave is deterministic \citep{Riles2023}. It involves contributions from the celestial motions of the Earth and the source, as well as the regular spin evolution of the source, e.g.~due to magnetic dipole braking \citep{Goldreich1969}. We assume that a time series $x(t)$ of interferometric detector data consists of a signal $h(t)$ and additive noise $n(t)$, where $t$ denotes the time at the detector. For the narrow frequency bands of interest for continuous gravitational waves, typically $\lesssim 1\,$Hz, we approximate $n(t)$ as white and Gaussian\footnote{Non-Gaussian features are present in real detector noise, often exhibiting ``line-like'' behavior, i.e.~at nearly constant frequency \citep{o23DetChar}. Distinguishing between these noise lines and astrophysical signals is not trivial, when there is no known instrumental source for the artefact \cite{Jaume2024}.}. In general, the time series $h(t)$ is a simple single-frequency sinusoid at the Solar System barycentre. It has amplitude modulations due to the time-varying antenna beam pattern of the detector(s), as well as Doppler modulations due to the revolution and rotation of the Earth. If the source is in a binary there is an additional Doppler modulation. Additional effects due to the Einstein delay, Shapiro delay, proper motion in the plane of the sky, and relativistic neutron star velocities are typically small, but are incorporated in most modern continuous gravitational wave search algorithms \citep{JKS98, Wette2023}. We elide the details of these additional effects in the discussion below. Explicitly, we write 
\begin{equation}
    h(t) = F_+(t)\, h_+(t) + F_\times(t)\, h_\times(t)\,,\label{eq:hoft_general}
\end{equation}
where $+$ and $\times$ refer to the plus and cross polarizations of the gravitational wave, and $F_A(t)$ is the detector response function to the $A$ polarization, defined in equations (10) and (11) in Ref.~\citep{JKS98}. The response functions are determined by the detector's location on Earth, and the polarization angle $\psi$ of the source. On the timescales of interest $\psi \in [-\pi/4,\pi/4]$ is a constant. 

For a source emitting at a single harmonic $f_{\rm GW} = 2f_{\rm rot}$, where $f_{\rm rot}$ is the rotation frequency of the neutron star (as would be the case for a perpendicular biaxial rotor), the source-dependent terms in Equation \eqref{eq:hoft_general} are
\begin{align}
    h_+(t) &= h_0 \frac{1 + \cos^2\iota}{2}\cos\left[\Phi(t) + \Phi_0\right]\,,\\
    h_\times(t) &= h_0 \cos\iota \sin\left[\Phi(t) + \Phi_0\right]\,,
\end{align}
where $h_0$ is the strain amplitude, $\iota$ is the inclination angle of the source, and $\Phi_0$ is an arbitrary phase offset. Again, on the timescales of interest we treat both $h_0$ and $\iota \in [0, \pi]$ as constant. 

The other component of the signal model is $\Phi(t)$, the phase of the gravitational wave as a function of the time at the detector. The standard (approximate) model of the phase is
\begin{align}
    \Phi(t) \approx &~2\pi \sum_{s=0}^k ~f_{\rm GW}^{(s)}(t_0) \frac{(t - t_0)^{s}}{s!} \left[ \frac{t - t_0}{s+1} 
     + \Delta_{R\odot}(t,\alpha,\delta) \nonumber \right. \\ 
     &~ + \Delta_{\rm B}(t, a_0, T_{\rm asc}, P)\big] \,,\label{eq:phioft_approx}
\end{align}
where the sum is typically truncated at $k=1$ or $k=2$, $t_0$ is the observation start time, and the superscript ${(s)}$ refers to the $s$-th time derivative. Both $\Delta_{R\odot}$ and $\Delta_{\rm B}$ are Rømer delays, arising from the Earth and source binary motion respectively. The projected semi-major axis is denoted by $a_0$, $P$ is the orbital period, $T_{\rm asc}$ is the time of passage through the ascending node (equivalent to the phase of the binary orbit), and $\alpha$ and $\delta$ are the right ascension and declination of the source. If the source is not in a binary, one has $\Delta_{\rm B}(t) = 0$. The detailed assumptions and approximations made in Equation~\eqref{eq:phioft_approx} are discussed at length in Refs. \citep{JKS98,Wette2023}, among other references.

To summarize, there are $10+k$ unknown parameters that define the canonical deterministic signal model for a neutron star in a binary emitting continuous gravitational waves at $f_{\rm GW} = 2f_{\rm rot}$. These parameters are $f_{\rm GW}(t=t_0)$, $\alpha$, $\delta$, $a_0$, $P$, $T_{\rm asc}$, $\Phi_0$, $\iota$, $\psi$, $h_0$, and the $k$ time derivatives of the frequency. Typically, searches for accreting binaries do not search over any secular frequency derivatives, as they are assumed to be too small to meaningfully shift the phase over the duration of a search \citep{o3amxp,o3vitsco,Whelan2023}. These searches also do not search over sky position, as the source location is known precisely from electromagnetic observations. The parameters $h_0$, $\Phi_0$, $\iota$, and $\psi$ are either set to their maximum likelihood estimates, or are marginalized over, depending on the search algorithm. This leaves four unknown parameters, over which template banks are constructed. The template banks are designed with a sufficient density of templates such that the maximum fractional loss in signal-to-noise ratio $\mu_{\rm max}$ (compared to a perfectly placed template on the exact signal parameters) is acceptable \citep{Leaci2015, Wagner2022,Mukherjee2023}. 

\subsection{Spin-wandering} \label{sec:sw_model}

\begin{figure*}
    \centering
    \includegraphics[width=0.8\linewidth]{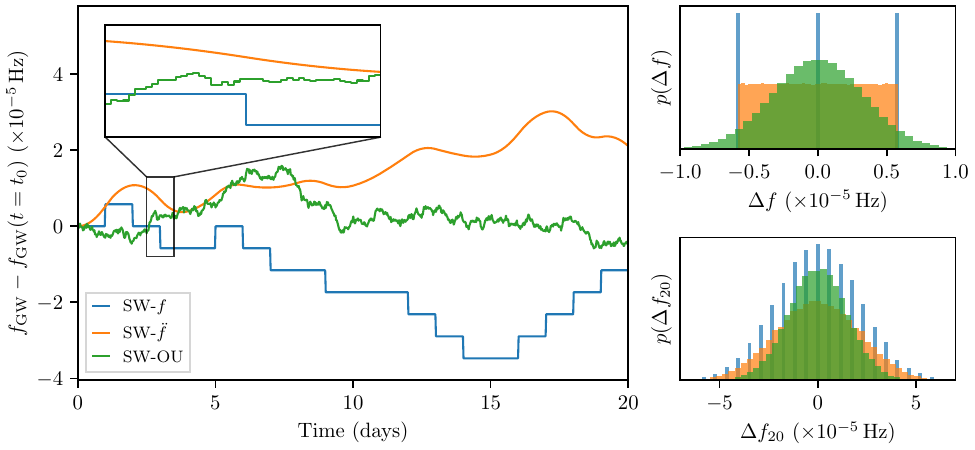}
    \caption{Left panel: Representative random realizations of $f_{\rm GW}(t)$ from three spin-wandering processes which can be injected into Gaussian noise: SW-$f$, SW-$\ddot{f}$, and SW-OU (in blue, orange, and green respectively). See text for details about these processes. Inset panel shows that both the SW-$f$ and SW-OU processes are composed of a piecewise-constant frequency evolution. Upper right panel: Histograms of $\Delta f = f_{\rm GW}(t + \tdr) - f_{\rm GW}(t)$ for the three processes (colors as in left panel). Lower right panel: Histograms of $\Delta f_{\rm 20} = f_{\rm GW}(t + 20 \tdr) - f_{\rm GW}(t)$ for the three processes (colors as in left panel). Histogram bar heights for the SW-$f$ process are adjusted such that the ordinate axes have the same scale. }
    \label{fig:compare_wander}
\end{figure*}

Some searches allow for stochasticity in the frequency (and hence phase) evolution, i.e. ``spin-wandering''. Spin-wandering may arise via a mechanism related to timing noise in young pulsars \citep{Cordes1985,Shannon2010}, or due to accretion torque variability in accreting systems \citep{Romanova2016,Mukherjee2018}. The exact physical cause is unknown. Replicating all of the observed properties of timing noise in pulsars within a phenomenological model is non-trivial. However, many pertinent features are preservered by modeling the system as a set of couple stcohastic differential (Langevin) equations, where the stochastic variations enter as white noise in $\ddot{f}_{\rm rot}$, resulting in red (i.e. frequency-dependent) noise in both the frequency and phase time-series. We refer the interested reader to Equations 2--5 of Ref.~\cite{Vargas2023bi} and references therein.

HMM searches explicitly include spin-wandering in the signal model by tracking the evolution of the gravitational wave frequency through a frequency--time trellis \citep{Suvorova2016,Suvorova2017,Melatos2021}. A HMM requires a transition matrix $A_{q_j q_i}$, which describes the probability that the hidden state moves from $q_i$ at time-step $t_n$ to $q_j$ at $t_{n+1}$. That is, it describes how much, how often, and in what manner the state is allowed to change between timesteps. Most searches track the hidden state $q(t_n) = f_{\rm GW}(t_n)$, i.e. the gravitational wave frequency after accounting for the Doppler modulations described in Section~\ref{sec:detphase}. Current implementations in the continuous gravitational wave context operate on regularly spaced data segments of length $1 \lesssim \tdr / (1\,\textrm{d}) \lesssim 30$, where $\tdr = t_{n+1} - t_n$ is kept constant. Historically, we set 
\begin{equation}
    A_{q_j q_i} = \frac{1}{3} \left(\delta_{q_j,q_{i+1}} + \delta_{q_j\,q_i} + \delta_{q_j,q_{i-1}} \right)\,, \label{eq:trans_third}
\end{equation}
where $\delta_{i,j}$ is the Kronecker delta. That is, the transition matrix allows $f_{\rm GW}$ to vary one frequency bin up or down (or stay in the same bin) every $\tdr$. Equation~\eqref{eq:trans_third} implicitly models $f_{\rm GW}$ as abruptly and discontinuously changing at the start of every segment. This model approximates the physical reality of a smoothly varying time series for $f_{\rm GW}$, with red noise statistics \citep{Cordes1985,Mukherjee2018,Goncharov2020,Serim2023}. One goal of Section~\ref{sec:det_eff} is to empirically quantify the impact on detection efficiency of using this idealized transition matrix when searching for signals generated with a different spin-wandering process.

A recent modification, which modestly improves sensitivity in certain regimes, also tracks the gravitational wave phase $\Phi(t_n)$ between segment boundaries \citep{Melatos2021}. This makes the hidden state two-dimensional, and requires an adjustment to the transition matrix and the HMM emission probabilities\footnote{By emission probability we refer to the likelihood that we observe the system in state $o_j$ given the hidden state is $q_i$; see section IIA of Ref.~\citep{Suvorova2016} for details.}, i.e. the $\mf$-statistic is replaced with a phase-dependent $\mathcal{B}$-statistic \citep{Melatos2021}. In particular, the alternative formulation assumes that $f_{\rm GW}$ undergoes a mean-reverting random walk, i.e. follows an Ornstein-Uhlenbeck process\footnote{A (well-parameterized) mean-reverting random process produces phase and frequency time-series that mimics observed timing noise in radio pulsars, see Refs.~\citep{Meyers2021,Antonelli2023,Vargas2023bi} for details.}, viz.
\begin{align}
    \frac{\td f_{\rm GW}}{\td t} &= -\gamma (f_{\rm GW} - \tilde{f}) + \sigma^2 \xi(t)\,,\label{eq:OUdamp}\\
    \frac{\td \Phi}{\td t} &= f_{\rm GW}\,, \label{eq:phidot}
\end{align}
where $\tilde{f}$ is the mean frequency towards which the process reverts, $\gamma^{-1}$ is the mean-reversion timescale, and $\sigma$ is the amplitude of fluctuations, which have white noise statistics: $\langle \xi(t) \rangle = 0$ and $\langle\xi(t)\xi(t')\rangle = \delta(t - t')$. Equations~\eqref{eq:OUdamp} and \eqref{eq:phidot} result in a transition matrix that is a $2\pi$-wrapped multivariate Gaussian \citep{Suvorova2018,Melatos2021}. 

\section{Injection procedure} \label{sec:inj}
\begin{table}
    \begin{ruledtabular}
    \caption{Fixed parameters used in the empirical tests in Section~\ref{sec:det_eff}. The start time $t_0$ is arbitrarily set to the start of O4. All other parameters are defined in Section~\ref{sec:signalmodel}.} \label{tab:static}
    \begin{tabular}{l l l}
        Parameter & Value & Unit \\
        \midrule
        $t_0$ & 1368921618 & GPS time \\
        $T_{\rm SFT}$ & 1800 & s \\
        $\tdr$ & 86400 & s \\
        $T_{\rm total}$ & 8640000 & s \\
        $\sqrt{S_X}$\footnotemark[1] & $5\times10^{-24}$ & Hz$^{-1/2}$ \\
        $f_{\rm GW}(t=t_0)$ & 234.56789 & Hz \\
        $\dot{f}_{\rm GW}(t=t_0)$ & 0 & Hz\,s$^{-1}$ \\
        $\ddot{f}_{\rm GW}(t=t_0)$ & 0 & Hz\,s$^{-2}$ \\
        $\alpha$ & 4.27569923844 & rad \\
        $\delta$ & $-$0.27297385834 & rad \\
        $\psi$ & $\pi/8$ & rad \\
        $\cos\iota$ & 1 & --- \\
        $a_0$\footnotemark[2] & 1.0 & lt-s \\
        $P$\footnotemark[2] & 432000 & s \\
        $T_{\rm asc}$ & 1373241618 & GPS time \\
    \end{tabular}
    \end{ruledtabular}
    \footnotetext[1]{We assume the same individual noise floor for two detectors located at LIGO Hanford and LIGO Livingston.}
    \footnotetext[2]{A wide binary with a five-day period and a projected semi-major axis of $1.0\,$lt-s allows all algorithms tested in Section~\ref{sec:det_eff} to use $T_{\rm SFT} = 1800\,$s without their linearising approximations breaking down. While low-mass X-ray binaries typically have orbital periods of less than one day~\cite{DiSalvo2022,Killestein2023}, binary pulsars are found with a wide range of periods, e.g. PSR J0437$-$4715 has a period of 5.74 days~\cite{Perera2019}. Of the 20 accreting millisecond X-ray pulsars searched in Ref.~\cite{o3amxp}, 18 have projected semi-major axes less than $1.0\,$lt-s, however many low-mass X-ray binaries, e.g. Sco X-1 and PSR J0437$-$4715, have a larger projected semi-major axis of $1.44\,$lt-s $\leq a_0 \leq 3.25\,$lt-s and $a_0 = 3.36672001(5)$\,lt-s respectively~\cite{Perera2019,Killestein2023}.}
\end{table}

Simulating spin-wandering signals using the tools implemented in the LVK Algorithm Library \citep{LAL2018} software \texttt{lalpulsar\_Makefakedata\_v5} (henceforth \texttt{MFD}) is not trivial\footnote{Other, potentially more flexible tools exist within LALSuite, such as the \texttt{simulateCW} Python module. We leave the integration of spin-wandering signals with these tools to future work.}. In this section we outline three possible procedures by which one may generate a synthetic spin-wandering continuous gravitational wave signal, and inject it into Gaussian noise using \texttt{MFD}\footnote{The performance of search algorithms in the presence of non-Gaussian noise, such as noise lines, and non-wandering signals is explored in Refs.~\citep{Keitel2015,Bayley2019,Kimpson2024}, among others. The performance of search algorithms in the presence of both non-Gaussian noise and spin-wandering signals is left to future work. This task is non-trivial, as whether the two stochastic elements are distinguishable is not obvious a priori. We note that the detection efficiency of search algorithms in the presence of noise lines also depends on the bespoke veto procedure adopted by any particular search.}. The \texttt{MFD} routine accepts constant values for $f_{\rm GW}$, $\dot{f}_{\rm GW}$ and $\ddot{f}_{\rm GW}$, as well as the other static parameters described in Section \ref{sec:detphase}. The routine injects the resultant signal into Gaussian noise with a specified amplitude spectral density $S_X$, or real data, yielding a frequency domain data product\footnote{The \texttt{MFD} routine within LALSuite can also produce time domain data, but time domain data are not analyzed in this paper.}. The length of each SFT $T_{\rm SFT}$ is fixed at 1800\,s in this paper, as that is a standard length produced by the LVK \citep{o3OpenData}\footnote{$\mf$-statistic-based algorithms use either a demodulation or resampling method to process the SFTs before matched filters are applied~\cite{Prix2009a,Singhal2019}. The computational trade-off that the CrossCorr algorithm need make between $T_{\rm SFT}$ and $T_{\rm max}$ is discussed in Ref.~\cite{Whelan2015}. }. We list the values of the static parameters used for the empirical tests in Section~\ref{sec:det_eff} in Table~\ref{tab:static}. We assume circular orbits in this paper for simplicity, although all search algorithms tested are able to search for signals that have non-zero eccentricity.

The simplest option, used to verify the functionality of the HMM-Viterbi algorithm in Refs.~\cite{Suvorova2016,Suvorova2017}, is to inject a signal with piece-wise constant $f_{\rm GW}$, which jumps discontinuously every $\tdr$. The statistics (i.e. allowed transitions) of such a process are defined by a probability density function (PDF) $p(\Delta f)$ with $\Delta f = f_{\rm GW}(t + \tdr) - f_{\rm GW}(t)$. We henceforth denote this spin-wandering option as ``SW-$f$''. 

A second option is to inject $\ddot{f}_{\rm GW}$ as a piece-wise constant function, which jumps every $\tdr$, and adjust $\dot{f}_{\rm GW}(t_n)$ and $f_{\rm GW}(t_n)$ such that $f_{\rm GW}$ is a smooth and continuous function. This is the methodology used in Ref.~\cite{Melatos2021}. Again, the choice of PDF, this time $p(\Delta \ddot{f})$ with $\Delta \ddot{f} = \ddot{f}(t_{n+1}) - \ddot{f}(t_n)$, determines the statistics of the process. We elaborate on the impact of various choices of $p(\Delta \ddot{f})$ in Appendix \ref{app:swddotf}. In summary, we find that $p[\Delta \ddot{f}(t_i)]$ for the $i$-th segment must be a function of $\dot{f}(t_{i-1})$ to ensure that the variance of the process does not grow with $i$. We henceforth denote this spin-wandering option as ``SW-$\ddot{f}$''.

The third option is to simulate an Ornstein-Uhlenbeck process directly, then sample the full frequency time series $f_{\rm GW}$ once per interval of duration $T_{\rm SFT}$, and treat the frequency as constant for the duration of each SFT. We henceforth denote this spin-wandering option as ``SW-OU''. The equations of motion for an Ornstein-Uhlenbeck process are presented in Equations~\eqref{eq:OUdamp} and \eqref{eq:phidot}. There are three free parameters: $\gamma$, $\sigma$, and $\tilde{f}$. We fix $\tilde{f} = f_{\rm GW}(t=t_0)$ in this paper. The impacts of $\gamma$ and $\sigma$ are discussed in Section~\ref{sec:swinj}.  

We show representative random realizations of the three options in the left panel of Figure~\ref{fig:compare_wander}. For each option (SW-$f$ in blue, SW-$\ddot{f}$ in orange, and SW-OU in green) we show 20\,days of the frequency evolution for a spin-wandering signal. Each of the options has respective parameters fixed such that the frequency moves $\sim 1/(2\tdr) \approx 5\times10^{-6}\,$Hz every $\tdr=1\,$day. In the top right panel we show the distribution of $\Delta f = f_{\rm GW}(t + \tdr) - f_{\rm GW}(t)$ for the three spin-wandering options. For SW-$f$ (blue histogram) the process can only move in discrete jumps of size $\pm 5\times10^{-6}\,$Hz. For SW-$\ddot{f}$ (orange histogram) the distribution $p(\Delta f)$ is uniform, given the choice for $p(\Delta \ddot{f})$ discussed in Appendix~\ref{app:swddotf}. For the SW-OU process (green histogram), a Gaussian is observed for $p(\Delta f)$, as expected for a process undergoing a random walk governed by Equations~\eqref{eq:OUdamp} and \eqref{eq:phidot}. By the central limit theorem, the long-term statistics of the three options converge, as shown in the bottom right panel of Figure~\ref{fig:compare_wander}; for example, the distribution $p(\Delta f_{20})$, with $\Delta f_{20} = f_{\rm GW}(t + 20\tdr) - f_{\rm GW}(t)$, approaches a Gaussian for each option. 

\section{Search procedure} \label{sec:alg}
Testing the detection efficiency of the full gamut of continuous gravitational wave search algorithms on spin-wandering signals is outside the scope of this paper. We restrict attention instead to four popular algorithms or approaches which have featured in several published continuous gravitational wave searches: 
\begin{enumerate*}[label=\roman*)]
    \item the fully coherent $\mf$-statistic \citep{JKS98,Cutler2005,Prix2007,Prix2009a},
    \item the semi-coherent $\mf$-statistic (sometimes called StackSlide) \citep{Brady2000,Prix2012},
    \item the CrossCorr algorithm \citep{crosscorInit,Whelan2015}, and
    \item the HMM-Viterbi algorithm using the $\mj$-statistic \citep{Suvorova2016,Suvorova2017}.
\end{enumerate*} 
We refer the reader to the respective methodology papers above for more on each of these search algorithms. We summarize the main points relevant to the tests in this paper in Sections~\ref{sec:fstat}--\ref{sec:hmm}.

All algorithms are provided the true injected values of $\alpha$, $\delta$, $a_0$, $P$, and $T_{\rm asc}$ for the tests in this paper. That is, we do not search a template bank for sky position or any binary orbital elements. We also fix $\dot{f}_{\rm GW} = 0$ and $\ddot{f}_{\rm GW}=0$ for each search algorithm. The three semi-coherent algorithms (semi-coherent $\mf$-statistic, CrossCorr, and HMM-Viterbi) have a frequency bin spacing of $\Delta f_{\rm bin} = 1 / (2 \tdr) \approx 5\times10^{-6}\,$Hz. For the fully-coherent $\mf$-statistic we set $\Delta f_{\rm bin} = 1 / (2 T_{\rm total}) \approx 5\times10^{-8}\,$Hz. We investigate the impact of $\Delta f_{\rm bin}$ on detecting spin-wandering signals in Appendix~\ref{app:delfbin}; it turns out that $\Delta f_{\rm bin}$ does not affect the sensitivity, so long as one has $\Delta f_{\rm bin} \lesssim 1 / (\tdr)$. The search band $\Delta f_{\rm band}$ is fixed to $\approx 0.0948\,$Hz\footnote{This search band is picked such that the number of frequency bins is a power of two, improving the computational efficiency of the $\mf$-statistic Fourier transform routines that run on Graphical Processing Units (GPUs) \citep{Dunn2022a}. We run the HMM-Viterbi search on GPUs.}, centered on $f_{\rm GW}(t=t_0)$.

\subsection{Fully coherent \texorpdfstring{$\mf$}{F}-statistic} \label{sec:fstat}
The fully coherent $\mf$-statistic constructs a matched filter for the deterministic phase model described by Equations~\eqref{eq:hoft_general}--\eqref{eq:phioft_approx} in Section~\ref{sec:detphase}, while fixing $\psi$, $\cos\iota$, $\Phi_0$, and $h_0$ to their maximum likelihood values. In this paper we use the resampling implementation in \texttt{LALPulsar\_ComputeFStatistic\_v2} \citep{Prix2009a, Patel2010}.

\subsection{Semi-coherent \texorpdfstring{$\mf$}{F}-statistic}
The semi-coherent $\mf$-statistic calculates the fully coherent $\mf$-statistic in successive segments, each of duration $\tdr < T_{\rm total}$. We then sum the $\mf$-statistic values across $N = T_{\rm total} / \tdr$ segments to produce one detection statistic at each frequency bin in the search band. This procedure implicitly models the continuous gravitational wave signal as having $f_{\rm GW} = {\rm constant}$, as we do not search any frequency derivatives, while allowing the phase to jump discontinuously between segments (c.f. equation~(79) of Ref.~\citep{Wette2023}). We use the resampling implementation of the $\mf$-statistic in \texttt{LALPulsar\_ComputeFStatistic\_v2} \citep{Prix2009a, Patel2010} to compute the $\mf$-statistic in the $N$ segments, combined with a simple Python wrapper to sum the computed values.

\subsection{CrossCorr}
The CrossCorr algorithm sums the cross-correlation of pairs of SFTs (from different detectors, or different times) with a filter which weights the pair according to the deterministic phase model described by Equations~\eqref{eq:hoft_general}--\eqref{eq:phioft_approx} in Section~\ref{sec:detphase}. One significant free parameter in this filter is $T_{\rm max}$, the maximum time allowed between SFT pairs. We set $T_{\rm max} = \tdr$. The inherent design of the CrossCorr algorithm makes it ``loosely-coherent''; the filter does not require the signal to be coherent across the entire duration $T_{\rm total}$, assuming $T_{\rm max} < T_{\rm total}$, but there are no boundaries in the search domain at which the phase is allowed to jump discontinuously. CrossCorr effectively marginalizes over $\psi$, $\cos\iota$, and $\Phi_0$, rather than setting them to their maximum likelihood values. The detection statistic that CrossCorr produces is an estimator of $h_0$.

\subsection{HMM-Viterbi with \texorpdfstring{$\mj$}{J}-statistic} \label{sec:hmm}
As described in Section~\ref{sec:sw_model}, the HMM-Viterbi algorithm explicitly allows $f_{\rm GW}$ to change over time. For the tests in this paper we use the transition matrix in Equation~\eqref{eq:trans_third}, as this is the most commonly adopted transition matrix in published continuous gravitational wave searches for accreting binaries \citep{Suvorova2017,o3amxp,o3vitsco}. This HMM allows both a frequency and a phase discontinuity between segments of duration $\tdr$. We use the $\mj$-statistic as the frequency domain estimator within each coherent segment. The $\mj$-statistic is a variant of the $\mf$-statistic which coherently sums the signal power dispersed into orbital sidebands. It offers computational savings as compared to the $\mf$-statistic when searching for continuous gravitational waves from a binary target. This efficiency is not needed in this work, as we do not search over a template bank of orbital parameters. However, we use the $\mj$-statistic for consistency with previous searches \citep{o3amxp,o3vitsco,o3vitscoredo}.

\section{Detection efficiency} \label{sec:det_eff}
\begin{figure*}
    \centering
    \includegraphics[width=0.9\linewidth]{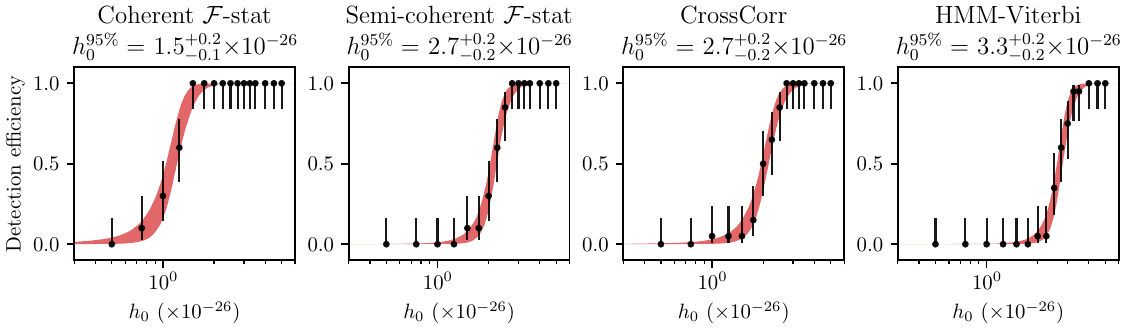}
    \caption{Detection efficiency of the coherent $\mf$-statistic, semi-coherent $\mf$-statistic, CrossCorr, and HMM-Viterbi algorithms when searching for a deterministic injected signal in Gaussian noise. Injection parameters are summarized in Table~\ref{tab:static}. Black points in each panel correspond to simple estimates of the efficiency via the ratio of recovered to injected signals at each value of $h_0$. Red bands in each panel correspond to the 95\% credible interval of the efficiency inferred via logistic regression. Panel titles include the median inferred $h_0$ at which the efficiency is equal to 95\%, with errors corresponding to the 95\% credible interval. }
    \label{fig:det_eff}
\end{figure*}

To determine detection efficiency, we first quantify what constitutes a ``detection''. Typically, continuous gravitational wave search algorithms mark a template as a candidate if the detection statistic $\rho$ exceeds a threshold $\rho_{\rm th}$ that depends on the probability of false alarm $p_{\rm FA}$. Thresholds may be calculated analytically \citep{JKS98}, if the distribution of the detection statistic in the presence of pure noise is known a priori. In practice, $\rho_{\rm th}$ is often estimated empirically, as real detector noise often makes the distribution of the detection statistic in the presence of pure noise vary across the search domain \citep{o3amxp,o3vitsco,Whelan2023}. To obviate calculating $\rho_{\rm th}$ for each algorithm, and to keep the detection criteria consistent across the different algorithms, in this paper we instead deem an injected signal to be ``detected'' if 
\begin{equation}
    \left| \bar{f}_{\rm GW} - f_{\rho,\,\textrm{max}} \right| \leq p_{\rm FA} \Delta f_{\rm band} \label{eq:simple_det}
\end{equation}
is satisfied, where $\bar{f}_{\rm GW}$ is the time-averaged frequency of the injected signal, and $f_{\rho,\,\textrm{max}}$ is the frequency of the highest detection statistic across the search band. The condition in Equation~\eqref{eq:simple_det} states that an algorithm that attempts to find an injected signal by randomly choosing a frequency within the search band would ``detect'' the signal a fraction $p_{\rm FA}$ of the time. For the HMM-Viterbi algorithm Equation~\eqref{eq:simple_det} is modified slightly: we replace the left-hand side with the time-averaged absolute difference between the injected frequency as a function of time, and the maximum likelihood frequency path, as recovered by the Viterbi algorithm. We remind the reader we provide the true injected values of $\alpha$, $\delta$, $a_0$, $P$, and $T_{\rm asc}$ to each algorithm, leaving only the gravitational wave frequency as the free parameter over which to search. We fix $p_{\rm FA} = 5\times10^{-4}$, but our results are broadly insensitive to this choice, provided $p_{\rm FA} \lesssim 10^{-2}$. The definition of detection in Equation~\eqref{eq:simple_det} is not useful for astronomical searches in real data, because we do not know the true frequency of the signal. However, it suffices for the empirical tests in this work where we know the true injected signal frequency.

In Section~\ref{sec:detinj} we estimate the detection efficiency of the four algorithms in the presence of a deterministic signal. In Section~\ref{sec:swinj} we estimate the detection efficiency (and hence the sensitivity depth) of the four algorithms in the presence of spin-wandering signals, with spin-wandering generated via three different procedures.

\subsection{Deterministic injections} \label{sec:detinj}
To calibrate expectations, we first calculate the detection efficiency of the four algorithms in the presence of a deterministic signal, i.e.~one which follows exactly the phase model described in Section~\ref{sec:detphase}. To do so we inject a signal with parameters as defined in Table~\ref{tab:static} into 20 realizations of Gaussian noise, at 16 values of $h_0$ between $0.5\times10^{-26}$ and $5\times10^{-26}$. We run the algorithms on each of the 320 injections and record whether the signal is detected via Equation~\eqref{eq:simple_det}. We then perform Bayesian logistic regression\footnote{All logistic regressions in this paper use the weakly informative priors suggested by Ref.~\citep{Gelman2008}.} \citep{Gelman2013} to estimate the detection efficiency as a function of $h_0$, viz.~$\varepsilon(h_0)$. 

\begin{figure*}
    \centering
    \includegraphics[width=0.99\linewidth]{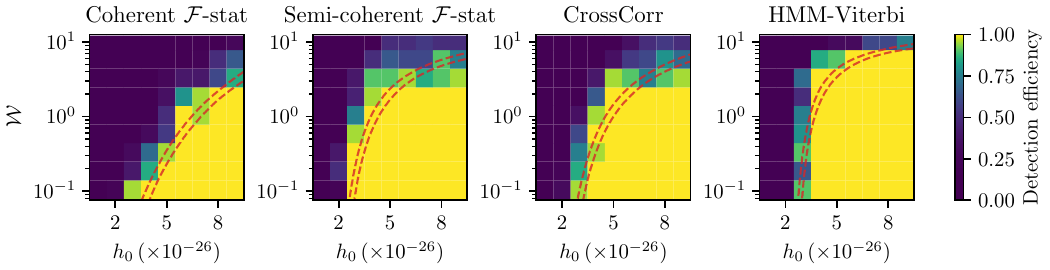}
    \caption{Detection efficiency of the coherent $\mf$-statistic, semi-coherent $\mf$-statistic, CrossCorr, and HMM-Viterbi algorithms when searching for a spin-wandering injected signal in Gaussian noise. The spin-wandering is injected as a SW-OU process, as described in Section \ref{sec:sw_model}. Static injection parameters are summarized in Table~\ref{tab:static}. The degree of spin-wandering is quantified by $\mathcal{W} = 2\Delta f_{\rm SW} \tdr$, i.e. when $\mathcal{W} = 1$ the frequency moves less than one semi-coherent frequency bin per $\tdr$. We fix $\gamma=10^{-12}\,$Hz. The color in each $\mathcal{W}$-$h_0$ pixel corresponds to the empirical detection efficiency, viz. the ratio of recovered to injected signals. Red dashed bands in each panel correspond to the 95\% credible interval of $h_0^{95\%}$ as inferred via multivariate logistic regression. }
    \label{fig:ou-lg}
\end{figure*}

Figure~\ref{fig:det_eff} shows the empirical efficiency (i.e. fraction of recovered injections) at each value of $h_0$ as black dots for each algorithm. The red band in each panel shows the 95\% credible interval of $\varepsilon(h_0)$. The inferred $h_0$ when the efficiency is 95\%, denoted by $h_0^{95\%}$, is $1.5^{+0.2}_{-0.1}\times10^{-26}$ for the coherent $\mf$-statistic, $2.7^{+0.2}_{-0.2}\times10^{-26}$ for both the semi-coherent $\mf$-statistic and CrossCorr algorithm, and $3.3^{+0.2}_{-0.2}\times10^{-26}$ for the HMM-Viterbi algorithm. We quote all values of $h_0^{95\%}$ with the central value as the median and error bars corresponding to the 95\% credible interval. Qualitatively, this ordering conforms with our expectations: the algorithm which performs the best is the one which assumes the exact signal model that is injected. The semi-coherent and loosely coherent algorithms perform slightly worse, and the HMM-Viterbi algorithm performs the worst of the four tested algorithms; its inherent signal model flexibility is redundant when searching for deterministic signals. The quantitative values of $h_0^{95\%}$, and the differences between the algorithms, depend on many factors, including but not limited to $T_{\rm total}$, the noise included in the injections, and $\tdr$ (for all algorithms besides the coherent $\mf$-statistic).

\subsection{Spin-wandering injections} \label{sec:swinj}
We now turn our attention to spin-wandering signals. We wish to investigate the detection efficiency of the four algorithms as a function of both $h_0$ and the degree of spin-wandering injected. To quantify spin-wandering for each of the processes outlined in Section~\ref{sec:sw_model}, we consider their individual distributions $p(\Delta f)$, with $\Delta f = f_{\rm GW}(t + \tdr) - f_{\rm GW}(t)$, as displayed in Figure~\ref{fig:compare_wander}. For SW-$f$ this distribution is discrete, with equal weight at $\Delta f = 0$ and $\Delta f = \pm \Delta f_{\rm SW}$, where $\Delta f_{\rm SW}$ is a free parameter which determines the size of frequency jumps. For SW-$\ddot{f}$ the shape of the distribution is controlled by $p(\Delta \ddot{f})$, as described in more detail in Appendix~\ref{app:swddotf}. As with SW-$f$, we can restrict the maximum frequency deviation over $\tdr$ to be at most $\Delta f_{\rm SW}$. For SW-OU, $p(\Delta f)$ depends predominantly on our choice of $\sigma$, so long as we have $\gamma \ll \tdr^{-1}$ \citep{Melatos2021}. By setting
\begin{equation}
    \sigma = \frac{\Delta f_{\rm SW}}{2\sqrt{\tdr}}\,,
\end{equation} 
we find $|\Delta f| < \Delta f_{\rm SW}$ 95\% of the time. We present a short justification for this in Appendix \ref{app:sigma_sw}. By defining 
\begin{equation}
    \mathcal{W} = 2\Delta f_{\rm SW}\tdr\,, \label{eq:wander_scale}
\end{equation}
we obtain a well-defined measure of the degree of injected spin-wandering. For example, for SW-$\ddot{f}$ or SW-$f$, when $\mathcal{W}=0.1$, 1, or 10, the gravitational wave frequency will not move more than 0.1, 1, or 10 frequency bin(s)\footnote{These frequency bins correspond to the standard frequency bin spacing for the HMM-Viterbi algorithms of $1/(2\tdr)$. We discuss alternative frequency bin spacings in Appendix~\ref{app:delfbin}.} every $\tdr$, respectively. For SW-OU the previous statement becomes probabilistic, i.e. the gravitational wave frequency will not move more than 0.1, 1, or 10 frequency bin(s) every $\tdr$, 95\% of the time.

\begin{figure*}
    \centering
    \includegraphics[width=0.99\linewidth]{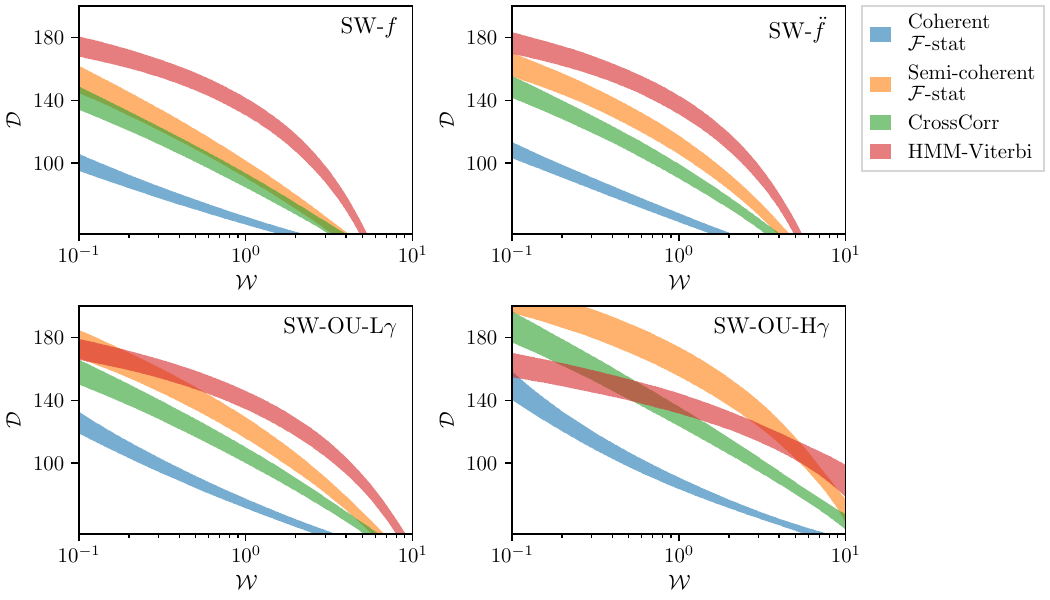}
    \caption{Sensitivity depth $\mathcal{D}$ of four different search algorithms when faced with four different varieties of spin-wandering signal injections. The spin-wandering processes are described in Section \ref{sec:sw_model}. Static injection parameters are summarized in Table~\ref{tab:static}. The degree of spin-wandering is quantified by $\mathcal{W} = 2\Delta f_{\rm SW} \tdr$, i.e. for $\mathcal{W} = 1$ the frequency moves less than one semi-coherent frequency bin per $\tdr$. SW-OU-L$\gamma$ (SW-OU-H$\gamma$) corresponds to the SW-OU process with $\gamma=10^{-12}\,$Hz ($\gamma=10^{-6}\,$Hz). Solid bands correspond to the 95\% credible interval of $\mathcal{D}$, as a function of $\mathcal{W}$, as inferred by multivariate logistic regression.}
    \label{fig:depth_ws}
\end{figure*}

We perform 20 injections into Gaussian noise at nine values of $\mathcal{W}$ logarithmically spaced between 0.1 and 10, and nine values of $h_0$ linearly spaced between $1\times10^{-26}$ and $9\times10^{-26}$. We record whether each algorithm can detect the injected signal using the criteria defined in Equation~\eqref{eq:simple_det}. This yields $9\times9\times20=1620$ binary values with which we perform Bayesian multivariate logistic regression \citep{Gelman2013}, where $\mathcal{W}$ and $h_0$ are the two variates. The regression includes a correlation term between $h_0$ and $\mathcal{W}$. That is, we model the detection efficiency as 
\begin{equation}
    \varepsilon(\mathcal{W}, h_0\,|\,\theta) = \frac{1}{1 + \exp\left(\beta_0 + \beta_1 h_0 + \beta_2 \mathcal{W} + \beta_3 h_0 \mathcal{W}\right)}\,,\label{eq:2deff}
\end{equation}
where $\theta = \{\beta_0, \beta_1, \beta_2, \beta_3\}$ is the set of parameters we infer through the logistic regression. Equation~\eqref{eq:2deff}, a simple two-dimensional sigmoid, is one choice for the likelihood function; one may opt to run the regression with a different likelihood, however the main results are unaffected.

We show the results for the SW-OU process, with a realistically low value of $\gamma=10^{-12}\,$Hz, in Figure~\ref{fig:ou-lg}. The color of each $\mathcal{W}$-$h_0$ pixel indicates the empirical detection efficiency, given the 20 injections at that pixel, with yellow indicating all 20 injections are recovered, and blue indicating that none of the 20 injections are recovered. The red dashed curves demarcate the 95\% credible interval of the two-dimensional curve $\varepsilon(h_0,\mathcal{W}) = 0.95$, i.e.~at a given value of $\mathcal{W}$ the two red dashed curves enclose the posterior credible estimate of the value of $h_0$ at which 95\% of the injected signals with that $\mathcal{W}$ are recovered, given the entire set of injections. For all the algorithms, as the degree of wandering increases, the value of $h_0$ at which the detection efficiency reaches 95\% increases. The semi-coherent algorithms (semi-coherent $\mf$-statistic, CrossCorr, and HMM-Viterbi) perform better than the fully coherent $\mf$-statistic even at low levels ($\mathcal{W}=0.1$) of spin-wandering. This is expected as the fully coherent algorithm applies a single matched filter across the entire search, while semi-coherent methods are more flexible, even if their inherent signal model does not explicitly include any stochastic frequency deviations. The HMM-Viterbi algorithm performs moderately better than the other algorithms for $\mathcal{W} \gtrsim 1$. As the spin-wandering decreases, with $\mathcal{W} \lesssim 1$, both the semi-coherent $\mf$-statistic and CrossCorr are slightly better at finding quiet injections, e.g. $h_0 = 3\times10^{-26}$. This reinforces the result noted in Section~\ref{sec:detinj}, that HMM-Viterbi performs relatively worse on deterministic signals. 

\begin{figure*}
    \centering
    \includegraphics[width=0.99\linewidth]{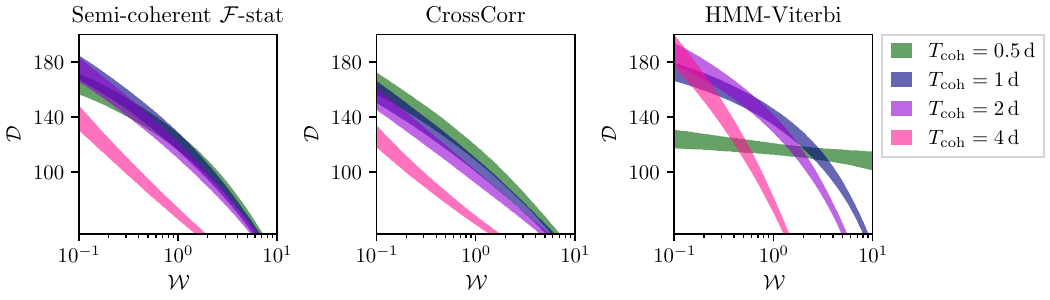}
    \caption{Sensitivity depth $\mathcal{D}$ of three semi-coherent search algorithms as a function of the degree of spin-wandering, with four different values of the coherent segment duration $\tdr$. The spin-wandering injected is generated via the SW-OU process with $\gamma=10^{-12}\,$Hz. Static injection parameters are summarized in Table~\ref{tab:static}. The degree of spin-wandering is quantified with $\mathcal{W} = 2\Delta f_{\rm SW} \tdr$, and $\tdr = 1\,$d is fixed for all injections, regardless of the $\tdr$ used to search for the injection. Solid bands correspond to the 95\% credible interval of $\mathcal{D}$, as a function of $\mathcal{W}$, as inferred by multivariate logistic regression.}
    \label{fig:tcoh_ws}
\end{figure*}
    
Figure~\ref{fig:depth_ws} summarizes the above results in terms of sensitivity depth for the suite of injections performed using SW-$f$, SW-$\ddot{f}$, SW-OU-L$\gamma$, and SW-OU-H$\gamma$ (the latter two processes correspond to SW-OU with $\gamma=10^{-12}\,$Hz and $\gamma=10^{-6}\,$Hz respectively). We plot the sensitivity depth\footnote{Note that we follow the definition of depth as in Equation (56) of Ref.~\cite{Wette2023}, in which the $S_X$ is normalised such that $\mathcal{D}$ is unitless.}
\begin{equation}
    \mathcal{D} = \frac{\sqrt{S_X / \textrm{Hz}}}{h_0^{95\%}}\, , \label{eq:depth}
\end{equation}
against $\mathcal{W}$, where $S_X$ denotes the single-sided power spectral density, and $h_0^{95\%}$ is the value of $h_0$ (at fixed $\mathcal{W}$) at which 95\% detection efficiency is reached, as inferred via logistic regression. We use $\mathcal{D}$ as a standard measure of the sensitivity of a search algorithm \citep{Behnke2015, Dreissigacker2018}, but note that it inherits the caveats already discussed with regard to comparing $h_0^{95\%}$ between different algorithms, i.e. its magnitude is conditional on the static parameters used in the injections (Table~\ref{tab:static}).  

We see in Figure~\ref{fig:depth_ws} that $\mathcal{D}$ decreases with $\mathcal{W}$ for all spin-wandering processes and search algorithms. Regardless of the spin-wandering process, the coherent $\mf$-statistic has the lowest sensitivity, if we have $\mathcal{W} \gtrsim 0.1$. (We reiterate that the coherent $\mf$-statistic has the greatest sensitivity for $\mathcal{W}=0$, as in Figure~\ref{fig:det_eff}.) For SW-$f$ and SW-$\ddot{f}$, HMM-Viterbi has the highest value of $\mathcal{D}$, provided we have $\mathcal{W}\gtrsim0.1$. This is expected, as it is the only search algorithm that explicitly includes spin-wandering in its signal model. The disparity is largest for SW-$f$, i.e. HMM-Viterbi performs best when we inject spin-wandering using the process assumed in its signal model. For SW-OU-L$\gamma$, HMM-Viterbi has similar depth to the semi-coherent $\mf$-statistic for $\mathcal{W}\lesssim 0.3$, but is slightly more sensitive at larger degrees of spin-wandering. For SW-OU-H$\gamma$, the short mean-reversion timescale results in the gravitational wave frequency not moving significantly during $T_{\rm obs}$, so the semi-coherent $\mf$-statistic and CrossCorr are more sensitive, as their signal model is more apt when the gravitational wave frequency does not move significantly over the search duration. We also perform a suite of injections with the SW-OU process and $\gamma=10^{-9}\,$Hz, but we do not show the results, as the inferred depths are statistically equivalent to the case with $\gamma=10^{-12}\,$Hz, i.e. the bottom-left panel of Figure~\ref{fig:depth_ws}.

It is important to note that searches using the semi-coherent $\mf$-statistic and CrossCorr typically follow-up candidates by increasing $T_{\rm coh}$ ($T_{\rm max}$ for CrossCorr). Figure~\ref{fig:tcoh_ws} shows that if the signal wanders, this follow-up will not increase detection confidence. For the semi-coherent $\mf$-statistic, the sensitivity gained by increasing the coherent integration time is initially offset by the sensitivity lost due to the signal wandering. However, by the time we increase $\tdr$ to $4\,$d we see that the evolution of $\mathcal{D}$ with $\mathcal{W}$ approaches that of the coherent $\mf$-statistic (c.f. the blue band in the bottom left panel of Figure~\ref{fig:depth_ws}). For CrossCorr, we see that moving from $\tdr=0.5\,$d to $\tdr=4\,$d reduces $\mathcal{D}$ at a fixed value of $\mathcal{W}$, i.e. the algorithm loses sensitivity with increasing coherence time, even when faced with signals with low values of spin-wandering $\mathcal{W} \approx 0.1$. We emphasize that even when $\mathcal{W} = 0.1$, i.e.~the stochastic daily variations in $f_{\rm GW}$ are within $5.7\times10^{-7}\,$Hz 95\% of the time, increasing the coherence time does not increase the sensitivity of the semi-coherent $\mf$-statistic nor CrossCorr algorithms.  An intuitive explanation is that as $T_{\rm coh}$ increases, the effective matched filter in frequency-space becomes more narrow for these algorithms. Therefore, the overlap between the stochastically-varying signal and the matched filter becomes smaller as $T_{\rm coh}$ grows. 

The depth of the HMM-Viterbi algorithm changes most dramatically with a changing $\tdr$, as the transition matrix changes with $\tdr$, changing the inherent signal model. When $\tdr=0.5\,$d the sensitivity decreases only slightly from $\mathcal{W}=0.1$ to $\mathcal{W}=10$, as the changing transition matrix in Equation~\eqref{eq:trans_third} allows for both more frequent and larger jumps in frequency. 

Choosing an appropriate $T_{\rm coh}$ for a given target and algorithm is a non-trivial task, as it depends on one's prior beliefs regarding the degree of spin-wandering present, and if it is present at all. If even a small amount of spin-wandering is present, increasing $T_{\rm coh}$ will eventually lead to a degradation of detection effeciency. Previous published searches (e.g. Refs.\cite{Whelan2023,o3vitscoredo,o3amxp}) have appealed to electromagnetic data (e.g. X-ray intensity fluctations) \cite{Mukherjee2018} and order-of-magnitude estimates \cite{Messenger2007,Sammut2014} to infer $T_{\rm coh}$ astrophysically for Sco X-1 and other low-mass X-ray binaries by analogy, but the $T_{\rm coh}$ values inferred thus are known to be uncertain. 

\section{Conclusion} \label{sec:concl}
Most continuous gravitational wave searches are undertaken without knowing precisely beforehand the degree to which the gravitational wave frequency wanders. Even when searches are guided by electromagnetic observations \citep{Mukherjee2018}, they are limited by the uncertainty regarding the underlying physical mechanism of spin-wandering. Allowing for spin-wandering in search algorithms increases the probability of detection in situations where reality does not produce signals that exactly match a canonical, deterministic matched filter \citep{Knee2023}. In this paper we systematically quantify the sensitivity of mature search algorithms to synthetic, software-injected spin-wandering signals. The efficacy of some of these algorithms when faced with deterministic synthetic signals was previously investigated in the Scorpius X-1 Mock Data Challenge \citep{scox1mdc1,Suvorova2017}.

We propose three different stochastic spin-wandering processes to produce realizations of a stochastically varying $f_{\rm GW}(t)$ which can be injected into Gaussian noise (or real data) using the standard tools within LALSuite \citep{LAL2018}. We perform suites of injections with different spin-wandering processes and magnitudes, which we quantify with $\mathcal{W} = 2\Delta f_{\rm SW} \tdr$, such that $\mathcal{W}=1$ corresponds to a spin-wandering process with typical variations in $f_{\rm GW}$ of less than $1/(2\tdr)$ (i.e.~the width of a frequency bin in a typical semi-coherent search) over a time-interval of duration $\tdr$. We attempt to detect these spin-wandering signals, of varying strain magnitudes, with four different algorithms: the fully coherent $\mf$-statistic \citep{JKS98}, the semi-coherent $\mf$-statistic \citep{Prix2012,Wette2018}, CrossCorr \citep{crosscorInit,Whelan2015}, and the HMM-Viterbi algorithm \citep{Suvorova2016,Suvorova2017}. Of the above algorithms, only the HMM-Viterbi algorithm explicitly includes spin-wandering in the signal model, by allowing the gravitational wave frequency to adjust between coherent segments.

The sensitivity depth $\mathcal{D}$ for deterministic ($\mathcal{D}_{\rm det}$) and spin-wandering ($\mathcal{D}_{\rm SW}$) signals depends both on the algorithm and spin-wandering process. As expected, we find that algorithms perform best when searching for signals that conform to their signal model. However, all algorithms retain some detection power when faced with spin-wandering signals. For example, with spin-wandering generated via an Ornstein-Uhlenbeck process with a long mean-reversion timescale (i.e. effectively a Wiener process) and $\mathcal{W}=1$, we find $\mathcal{D}_{\rm det} / \mathcal{D}_{\rm SW}=4.39^{+0.23}_{-0.27}$, $1.51^{+0.02}_{-0.03}$, $1.75^{+0.04}_{-0.04}$, and $1.07^{+0.01}_{-0.02}$ for the coherent $\mf$-statistic, semi-coherent $\mf$-statistic, CrossCorr, and HMM-Viterbi algorithms respectively. 

We warn against search algorithms attempting to increase detection significance of first-pass candidates by increasing the duration over which the signal is coherently integrated, if there is reason to believe the spin wanders. By increasing the coherence time, the sensitivity depth approaches that of the coherent matched filter, which does not allow for spin wandering. As a result the sensitivity depth decreases, the opposite of what is desired. Quantitatively, we see that $\mathcal{D}$ for the semi-coherent algorithms drops by a factor of $\sim2$ when we increase the coherence from $\tdr=1\,$d to $\tdr=4\,$d, when the injected signal is wandering with $\mathcal{W}=1$.

We reiterate that strain upper limits calculated after a null detection in a given search are conditional on the assumed signal model. Upper limits inferred via suites of injections of canonical, deterministic signals are not equivalent to upper limits calculated via suites of injections of spin-wandering signals. This important point has been emphasized in several published searches \citep{Beniwal2021,o3aSNR,o3vitsco,o3amxp,o3vitscoredo}. It would be beneficial, when computationally and scientifically feasible, for future searches involving real data to use both spin-wandering and canonical injections when setting upper limits, and to report both of upper limits. This would be particularly valuable in cases where there is reason to believe that spin-wandering may be present in the search target.

\section*{Acknowledgments}
This work was supported by the Australian Research Council (ARC) via the ARC Centre of Excellence for Gravitational Wave Discovery, Grant Number CE170100004. This work used computational resources of the OzSTAR national facility at Swinburne University of Technology. OzSTAR is partly funded by Swinburne University of Technology and the National Collaborative Research Infrastructure Strategy (NCRIS). This material is based upon work supported by NSF’s LIGO Laboratory which is a major facility fully funded by the National Science Foundation. This work has been assigned LIGO document number P2400149.

\emph{Software:}  \texttt{LSC Algorithm Library} \citep{LAL2018}, \texttt{Numpy} \citep{Harris2020}, \texttt{Scipy} \citep{Virtanen2020}, \texttt{Matplotlib} \citep{Hunter2007}, and \texttt{Stan} \citep{stan2022} through the \texttt{cmdstanpy} interface. The simple Python scripts that produce the spin-wandering simulations described in Section~\ref{sec:inj} are available at the following code repository: \url{https://git.ligo.org/unimelb/spinwandering_injector}.

\appendix
\section{Generating a spin-wandering signal with the \texorpdfstring{SW-$\ddot{f}$}{SW-ddotf} process}\label{app:swddotf}
To construct a spin-wandering signal which is smoothly continuous in $f$ and $\dot{f}$, but has step-wise jumps in $\ddot{f}$ at times $t_i$, with $\tdr = t_i - t_{i-1}$, at each time $t_i$ we pick a new value of $\ddot{f}(t_i)$, then adjust 
\begin{align}
    f(t_i) &= f(t_{i-1}) + \dot{f}(t_{i-1}) \tdr + \frac{1}{2}\ddot{f}(t_i)\tdr^2\,, \label{eq:f_cont} \\ 
    \dot{f}(t_i) &= \ddot{f}(t_i) \tdr\,.\label{eq:fdot_cont}    
\end{align}
After $N$ segments, we may recursively expand $\dot{f}_0^{(N)}$ and collect like terms to write
\begin{equation}
    f(t_N) = f(t_0) + \frac{\tdr^2}{2} \sum_{j=1}^{N} \left(2j - 1\right)\ddot{f}(t_j)\,,\label{eq:ftn_ddotf}
\end{equation}
where we assume $\dot{f}(t_0) = 0$. If we assume $\ddot{f}(t_j)$ for all $1 \leq j \leq N$ is drawn from the same distribution with variance $\zeta$, we see that the variance of Equation~\eqref{eq:ftn_ddotf} grows as 
\begin{align}
    \textrm{Var}\left[f(t_N)\right] &= \frac{\tdr^4}{4} \sum_{j=1}^{N} \left(2j - 1\right)^2 \zeta \label{eq:a4}\\
    &= \frac{\tdr^4 \zeta N}{12} \left(4N^2 - 1\right)\,, \label{eq:a5}
\end{align}
that is, the variance scales like $N^3$ for $N\gg1$. The step from Equation~\eqref{eq:a4} to \eqref{eq:a5} is a purely algebraic manoeuvre. This implies that the variance of $\Delta f(t_N) = f(t_N) - f(t_{N-1})$ grows with time, viz.
\begin{equation}
    \textrm{Var}\left[\Delta f(t_N)\right] = \frac{\tdr^3 \zeta N(N-1)}{4}\,. \label{eq:varddotf}
\end{equation}
Equation~\eqref{eq:varddotf} is unphysical; the variance of the difference in frequency from one coherent segment to the next should not depend on the number of segments $N$. Ref.~\citep{Melatos2021} circumvented the issue by drawing a new value of $\ddot{f}(t_i)$, whenever the proposed value yields $|f(t_{i+1}) - f(t_{i})| > \Delta f_{\rm SW}$, the maximum desired frequency deviation between two segments. Hence the PDF $p[\Delta \ddot{f}(t_i)]$ is not constant; it is a function of $\dot{f}(t_{i})$. 

Explicitly, when simulating a spin-wandering signal with the SW-$\ddot{f}$ process, we choose
\begin{equation}
    p\left[\ddot{f}(t_i)\right] \sim \frac{2}{\tdr^2}\, \mathcal{U} \left\{-\Delta f_{\rm SW},\, \left[\Delta f_{\rm SW} - \dot{f}(t_{i}) \tdr\right]\right\} \label{eq:pddotfa}
\end{equation}
for $\dot{f}(t_i) > 0$ and
\begin{equation}
    p\left[\ddot{f}(t_i)\right] \sim \frac{2}{\tdr^2}\, \mathcal{U} \left\{-\left[\Delta f_{\rm SW} + \dot{f}(t_{i}) \tdr\right],\, \Delta f_{\rm SW} \right\} \label{eq:pddotfb}
\end{equation} 
for $\dot{f}(t_i) < 0$, where we write $X \sim \mathcal{U}(a, b)$ if $X$ is a random variable drawn from a uniform distribution bounded below by $a$ and above by $b$. After picking a suitable value for $\ddot{f}(t_i)$ from Equation~\eqref{eq:pddotfa} or \eqref{eq:pddotfb} we update $f(t_i)$ and $\dot{f}(t_i)$ via Equations~\eqref{eq:f_cont} and \eqref{eq:fdot_cont} respectively. We also ensure that the gravitational wave phase is continuous at segment boundaries $t_i$. Generating the SW-$\ddot{f}$ process with the above choice for $p[\ddot{f}(t_i)]$ produces a uniform distribution for $p(\Delta f)$ between $-\Delta f_{\rm SW}$ and $\Delta f_{\rm SW}$, as displayed in the orange histogram in the top-right panel of Figure~\ref{fig:compare_wander}.

\section{Relationship between \texorpdfstring{$\sigma$}{sigma} and \texorpdfstring{$\Delta f_{\rm SW}$}{Deltafsw} for the SW-OU process} \label{app:sigma_sw}
The Ornstein-Uhlenbeck process leads to a diffusion equation; see section IIIB and appendix C of Ref.~\citep{Melatos2021}. The solution for $f(t_i)$ conditioned on the starting frequency $f(t_0)$ is a Gaussian process with mean \citep{Gardiner2009}
\begin{align}
    \mathbb{E}[f(t_i)\,|\,f(t_0)] = &~f(t_0)\exp[-\gamma (t_i - t_0)] \nonumber \\
    &+ \tilde{f} \{1 - \exp[-\gamma (t_i - t_0)]\}\,,
\end{align}
and covariance 
\begin{align}
    \textrm{cov}[f(t_i), f(t_0)] =&~ \frac{\sigma^2}{2\gamma} \big\{ \exp[-\gamma (t_i - t_0)] \nonumber \\
    &+ \exp[-\gamma (t_i + t_0)]\big\}\,.
\end{align}
If we want differences $\Delta f = f(t_i + \tdr)  - f(t_i)$ to be no larger in magnitude than $\Delta f_{\rm SW}$, 95\% of the time (i.e. two standard deviations of the Gaussian process), we require
\begin{align}
    \big|\mathbb{E}[f(t_i + \tdr)\,|\,f(t_i)] &+ 2\sqrt{\textrm{cov}[f(t_i + \tdr), f(t_i)]} \big| \nonumber \\
    &\approx \Delta f_{\rm SW}\,, \label{eq:full_driftcov}
\end{align}
Solving Equation~\eqref{eq:full_driftcov} to find $\sigma$ as a function of $\gamma$, $\tdr$, $\tilde{f}$, $\Delta f_{\rm SW}$, and $f(t_i)$ is left to future work. We find empirically that it suffices to fix
\begin{equation}
    \sigma = \frac{\Delta f_{\rm SW}}{2 \sqrt{\tdr}}\,, \label{eq:approx_sig2}
\end{equation}
assuming $\gamma \tdr \ll 1$, and $|\tilde{f} - f(t_i)| < \gamma \tdr$. Both conditions hold for all simulations in this paper.

\section{Impact of frequency bin spacing} \label{app:delfbin}
\begin{figure*}
    \centering
    \includegraphics[width=0.9\linewidth]{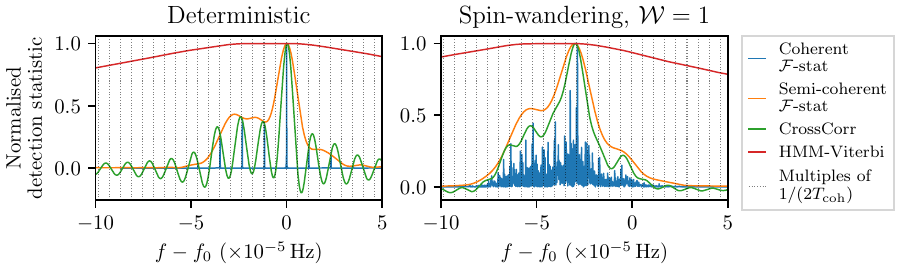}
    \caption{Behaviour of the peak-normalised detection statistics as a function of frequency for four search algorithms (curves color-coded as per the legend) in the vicinity of a loud injection ($h_0 = 10^{-23}$, with other static parameters including $f_0=f_{\rm GW}(t=t_0)$ listed in Table~\ref{tab:static}) for both a deterministic (left panel) and spin-wandering (right panel) signal. The spin-wandering signal is generated via the SW-OU process, with $\mathcal{W}=1$ and $\gamma=10^{-12}\,$Hz. Multiples of $1/(2\tdr)$ are shown as dotted black vertical lines. These lines mark the spacing of frequency bins used in this paper for the semi-coherent $\mf$-statistic, CrossCorr, and HMM-Viterbi algorithms. We see that the spin-wandering injection has signal power spread out over a wider band in frequency compared to the canonical deterministic injection, as expected.}
    \label{fig:freq_mismatch}
\end{figure*}

Most searches using the HMM-Viterbi algorithm set the frequency bin spacing to $\Delta f_{\rm bin} = 1 / (2\, \tdr)$ \citep{o3aSNR,o3amxp,o3vitsco}. Other searches typically set their frequency bin spacings with reference to a parameter-space metric, and a desired maximum mismatch $\mu_{\rm max}$. The maximum mismatch is defined as the fractional loss in squared signal-to-noise ratio when searching for the signal using the closest-matching template in the template bank compared to the template which perfectly matches the signal. A first-order expansion of the metric, assuming all search parameters besides frequency are known, gives 
\begin{equation}
    \Delta f_{\rm bin} = \frac{\sqrt{12\mu_{\rm max}}}{\pi \tdr}\,,
\end{equation}
from Equation (11) of Ref.~\citep{Wette2008}, which suggests one should set $\Delta f_{\rm bin} \approx 1 / (2\,\tdr)$ for $\mu_{\rm max} \approx 0.2$, for example. The above estimate is refined in Ref.~\cite{Wette2015}, which derives the spacing for the complete parameter space of a semi-coherent $\mf$-statistic search. 

Suppose one performs a discrete Fourier transform on data spanning a time $T$. If the spacing of Fourier bins is $1 / T$ the complex amplitudes of adjacent frequency bins are uncorrelated if the data contains Gaussian noise \citep{Ransom2002}. Over-resolution is achieved in the continuous gravitational wave search context by setting $\Delta f_{\rm bin} < 1 / \tdr$, and reduces the impact of spectral leakage on finding the frequency bin with the most power (although see Refs.~\cite{Thomson1982} and \cite{Percival1993} for alternative harmonic detection and estimation approaches). The information present in the data does not increase as $\Delta f_{\rm bin}$ decreases. However, finer frequency bin spacing may reduce the probability of false dismissal in a low signal-to-noise regime. 

The mismatch in frequency is inherently hard to calculate for spin-wandering signals, as a stochastic matched filter that perfectly matches the frequency evolution of the signal cannot be ingested by the search algorithms in use at the time of writing \citep{Liu2023}. One way to visualize the response of search algorithms in the frequency domain is to inject a strong signal (e.g. $h_0=10^{-23}$, with all other parameters as listed in Table~\ref{tab:static}) and run the algorithms with a fine frequency spacing near the injection. We show the results in Figure~\ref{fig:freq_mismatch}, for a deterministic and a spin-wandering injection. All algorithms have a peak in their detection statistic at the same frequency $f_{\rm p}$, as expected for such a loud injection. However, the detection statistic falls away with $|f-f_{\rm p}|$ differently for different algorithms, as described below.

\begin{figure*}
    \centering
    \includegraphics[width=0.99\linewidth]{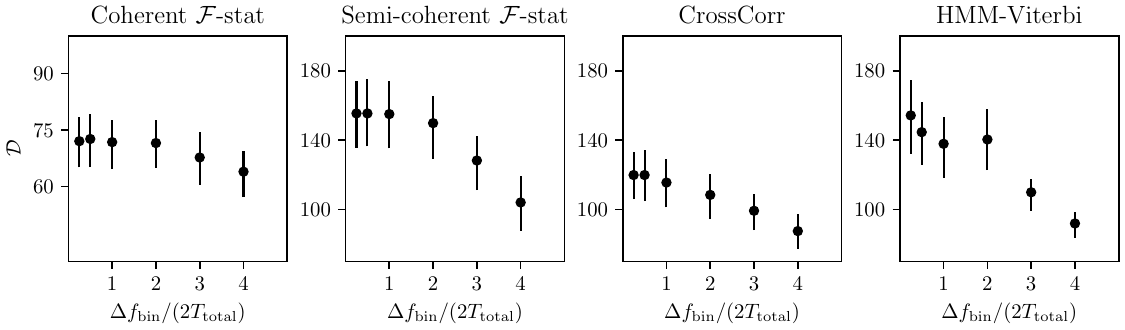}
    \caption{Sensitivity depth $\mathcal{D}$ in the presence of a spin-wandering signal as a function of frequency bin spacing $\Delta f_{\rm bin}$ for four different search algorithms. The spin-wandering process is SW-OU-L$\gamma$, with $\mathcal{W}=1$. Error bars span the 95\% credible interval of $\mathcal{D}$.}
    \label{fig:delfbin_depth}
\end{figure*}

For the deterministic injection, the coherent $\mf$-statistic collects almost all power within very narrow peaks, spaced $(1\,\textrm{d})^{-1}$ apart. The highest peak of the detection statistic has a full-width half-maximum (FWHM) of $\sim 10^{-7}\,$Hz $\approx 1/T_{\rm total}$. For the semi-coherent $\mf$-statistic (CrossCorr) the FWHM of the detection statistic is $\sim10^{-5}\,$Hz $\approx 1/\tdr$ [$\sim6\times10^{-6}\,$Hz $\approx 1/(2\tdr)$]. The FWHM for the HMM-Viterbi algorithm is broad, at $\sim4\times10^{-4}\,$Hz $\approx 100 / (2\tdr)$, because bins near the injected signal are highly correlated; many paths through the frequency-time trellis include the strong injected signal. While it may appear there are two peaks for the HMM-Viterbi algorithm, this is a visual artefact, the algorithm recovers the highest detection statistic at the frequency at which the signal is injected.

For the spin-wandering injection the power of the coherent $\mf$-statistic in the frequency domain is dispersed into many peaks due to the wandering gravitational wave frequency; the FWHM of the central peak is approximately 1.5 times larger than for the deterministic signal. The FWHMs of the peaks for the semi-coherent $\mf$-statistic and CrossCorr algorithms are both approximately three times larger than for the deterministic signal. The FWHM of the HMM-Viterbi algorithm in the presence of the spin-wandering signal is the same as when in the presence of a deterministic signal, as expected. As $\mathcal{W}$ increases, the FWHM of the coherent $\mf$-statistic, semi-coherent $\mf$-statistic, and the CrossCorr algorithms also increases, as the gravitational wave frequency wanders through more of the search band. 

For deterministic signals, empirical studies show that semi-coherent algorithms have $\mathcal{D} \propto \tdr^{w/4}$, with $1\leq w \leq 2$ (for fixed $T_{\rm total}$ and noise floor) \citep{Prix2012, Wette2012}. When faced with a potentially spin-wandering signal, the accepted rule-of-thumb is that one should set $\tdr$ as long as possible, such that the signal does not move more than one frequency bin, to achieve the maximum sensitivity depth. This implies that we should set $\Delta f_{\rm bin}$ as large as feasible, while ensuring that $\Delta f_{\rm bin}$ is less than the FWHM expected for the search algorithm in the presence of a spin-wandering signal.  

Can we set $\Delta f_{\rm bin} = 1 / \tdr$ as opposed to $1 / (2\tdr)$ or smaller when searching for spin-wandering signals without losing sensitivity? Figure~\ref{fig:delfbin_depth} appears to answer this in the affirmative. We see that $\mathcal{D}$ decreases for $\Delta f_{\rm bin} \geq 3/(2\tdr)$ for all algorithms, but is constant (within the 95\% credible interval) for smaller values of $\Delta f_{\rm bin}$. The injections used in Figure~\ref{fig:delfbin_depth} generate spin-wandering via the SW-OU process, with $\gamma=10^{-12}\,$Hz and $\mathcal{W}=1$. Static injection parameters are listed in Table~\ref{tab:static}, except that the frequency band size is increased to $\Delta f_{\rm band} \approx 0.379\,$Hz. The increase in $\Delta f_{\rm band}$ maintains the efficacy of the detection criterion in Equation~\eqref{eq:simple_det} even with large frequency bin spacings (i.e. fewer bins in the search band).  In summary, we recommend that search algorithms do not over-sample in the frequency domain, as it does not significantly impact the sensitivity depth of the search, when a spin-wandering signal is present.

As a worked example, the HMM-Viterbi algorithm searched for continuous gravitational waves from Scorpius X-1 in LIGO O2 and O3 data. Both searches adopted $\tdr = 10\,$d \citep{o2vitsco,o3vitsco}, and the standard frequency bin spacing $1/(2\,\tdr) \approx 5.8\times10^{-7}\,$Hz. We could instead choose $\Delta f_{\rm bin} = 1/\tdr \approx 1.2\times10^{-6}\,$Hz, or $\Delta f_{\rm bin} = 1 / (4\, \tdr) \approx 2.9\times10^{-7}\,$Hz. The HMM transition matrix defined in Equation~\eqref{eq:trans_third} allows the gravitational wave frequency to change by up to one frequency bin every $\tdr$. Hence, over the course of a year (i.e. 36 segments of length $\tdr$), the default choice allows the frequency to move a maximum of $2.1\times10^{-5}\,$Hz, while the latter two choices correspond to maximum changes in frequency of $4.2\times10^{-5}\,$Hz and $1.0\times10^{-5}\,$Hz respectively. If we believe the maximum total frequency deviation is below $\approx 5\times10^{-5}$\,Hz over one year \citep{Mukherjee2018}, we can keep $\Delta f_{\rm bin} = 5.8\times10^{-7}\,$Hz but set $\tdr = 20\,$d, potentially increasing the sensitivity by up to 16\% (via the heuristic $\mathcal{D} \propto \tdr^{1/4}$ for deterministic signals). However, we remind the reader that, according to Figure~\ref{fig:tcoh_ws}, increasing $\tdr$ does not necessarily increase sensitivity, if the signal is spin-wandering.

\bibliography{spinwander}

\end{document}